\documentstyle[12pt]{article}
\begin{document}
\title{\bf More on triangular mass matrices 
for fermions } 
\author{ H. B. Benaoum \\
Institut f\"ur Physik,  Theoretische Elementarteilchenphysik, \\ 
Johannes Gutenberg--Universit\"at,  55099 Mainz, Germany \\
email : benaoum@thep.physik.uni-mainz.de
 }
\date{ }
\maketitle
~\\
\abstract{ A direct proof is given here which shows that instead of 6 complex 
numbers, the triangular mass matrix for each sector could just be    
expressed in terms of 5  by performing a specific weak basis transformation, 
leading therefore to a new textures for triangular mass matrices. \\ 
Furthermore, starting with the set of 20 real parameters for both sectors, 
it can shown that 6 redundant parameters can be removed by using the 
rephasing freedom.
} \\
~\\
~\\
~\\
{\bf MZ--TH/98--37 } \\
\newpage
~\\
One of the major unsolved problem in particle physics is to understand 
flavor mixing and fermion masses, which are free parameters in the 
standard model. \\
Many speculations have been made and some Ans\"atze were proposed by 
introducing extra symmetries to cast the mass matrices in some particular 
forms~\cite{fri,wil,geo,ram}. \\
A popular Ansatz suggested by Fritzsch is the Nearest--Neighbor 
Interactions ( NNI )~\cite{fri} which has been for a long time a mysterious 
parametrization. Branco et al.~\cite{bra} have been the first to show that 
for non--hermitian mass matrices some textures \`a la Fritzsch were just 
a rewriting of the mass matrices in a special basis without any loss of 
generality by using this basis. \\
~\\
Making use of the freedom in choosing the right--handed bases for the fermion 
fields, a class of triangular mass matrices has been introduced recently. 
It was shown, in particular, that the effective triangular mass matrix 
${\cal T}_{eff}$, which is obtained by shifting the diagonalization in one 
sector, is reconstructed analytically from the moduli of the CKM matrix and 
quark masses. This reconstruction is unique up to trivial phase 
redefinitions. \\
Such patterns arise in the framework of 
Marseille--Mainz noncommutative geometry~\cite{coq}, namely triangular mass 
matrices~\cite{hau,sch}. They are typical for reducible but indecomposable 
representations of graded Lie algebras,
\begin{eqnarray}
\left( \begin{array}{ccc}
t_{11} & 0 & 0 \\
t_{21} & t_{22} & 0 \\
t_{31} & t_{32} & t_{33} \end{array} \right)~~~~~,~~
\left( \begin{array}{ccc}
t_{11} & t_{12} & t_{13} \\
0 & t_{22} & t_{23} \\
0 & 0 & t_{33} \end{array} \right)
\end{eqnarray}
where $t_{11},t_{21},t_{22},t_{31},t_{32},t_{33}$ for lower--triangular 
mass matrix ( resp. $t_{11},t_{12},t_{13},t_{22},t_{23},t_{33}$ for 
upper--triangular mass matrix ) are 6 complex numbers. Of    
course, not all of these parameters are physically relevant, since some of 
them can be made real by performing a suitable phase transformation on the 
fermion fields.  \\
In a recent paper~\cite{ben}, see also~\cite{sss},  
an indirect proof was given that there is   
a weak basis transformation that reduces the number of complex  
parameters to 5 for each sector. We have 
also made a bridge and a close connection between the NNI and 
triangular mass matrices.  \\
~\\
Here, we present a direct proof to express these triangular mass matrices in 
a economic and concise way with 5 parameters, through a specific weak basis 
transformation which mean that the extra parameter is either zero or 
dependent of the others. \\
In fact, it can be shown that from the set of 20 real parameters for both 
sectors, 6 phases can be removed by exploiting the freedom in redefining the 
fermion fields leaving therefore 5 real moduli and two real phases for each 
sector. Moreover one sector can be made completely real with 5 real moduli 
and consequently the other sector has 5 real moduli and two real relative 
phases. \\
~\\
It is well known, within the standard model that the two sets of mass  
matrices $( M_u, M_d )$ and 
$( {\cal M}_u, {\cal M}_d )$ related to each other through, 
\begin{eqnarray}
M_u~~=~~U^{\dagger} {\cal M}_u V_u ,~~~M_d~~=~~U^{\dagger} {\cal M}_d V_d 
\end{eqnarray}
give rise to the same physics ( i.e. same masses and mixings ). \\
We have then the following two propositions for trianguar mass matrices, \\
~\\
\underline{ Theorem1 :} ( for two lower triangular mass matrices ) \\ 
Given any two nonsingular $3 \times 3$ lower triangular mass matrices 
${\cal T}_u, {\cal T}_d$, there exists always a weak 
basis transformation such that the 
new mass matrices $T_u, T_d$ are lower triangular with vanishing matrix 
elements (2,1). \\
~\\
\underline{ Proof :} \\
we first prove that the transformed mass matrices are lower triangular. \\
Indeed, ${\cal T}_{u,d}$ transform as 
$U^{\dagger} {\cal T}_{u,d} V_{u,d}$. It is clear 
that, for a fixed unitary $U$, the unitary matrices $V_{u,d}$ could always be 
chosen such that $T_{u,d} = U^{\dagger} {\cal T}_{u,d} V_{u,d}$ are lower 
triangular. \\
To prove the second part, i.e $( T_{u,d} ){_{21}} = 0$, just consider  
the hermitian mass matrices ${\cal H}_u = {\cal T}_u {\cal T}_u^{\dagger}$    
and ${\cal H}_d = {\cal T}_d {\cal T}_d^{\dagger}$ which transform as 
$H_{u,d} = U^{\dagger} {\cal H}_{u,d} U$ with :
\begin{eqnarray} 
( H_u ){_{21}} & = & ( H_d ){_{21}}~~=~~ 0. 
\end{eqnarray}
and construct the unitary matrix $U$ such that :
\begin{eqnarray}
U^{\star}_{i2} ({\cal H}_u){_{ij}} U_{j1} & = & 0 \nonumber \\
U^{\star}_{i2} ({\cal H}_d){_{ij}} U_{j1} & = & 0 \nonumber \\ 
U^{\star}_{i2} U_{i1} & = & 0
\end{eqnarray}
which means that the vector $U^{\star}_{i2}$ is orthogonal to the three 
vectors $({\cal H}_u){_{ij}} U_{j1}, 
({\cal H}_d){_{ij}} U_{j1}$ and $U_{i1}$. \\
For this to be possible, choose $U_{i1}$ such that these three vectors are 
linearly dependent, i.e. 
\begin{eqnarray}
a U_{i1} + b ({\cal H}_u){_{ij}} U_{j1} + c ({\cal H}_d){_{ij}} U_{j1} & = & 0
\end{eqnarray}
where $a,b$ and $c$ are some nonvanishing parameters. \\
Therefore $U_{i1}$ is just a normalized eigenvector of the matrix 
${\cal H}_u + \frac{c}{b} {\cal H}_d$, with eigenvalue $- \frac{a}{b}$. \\
Starting from $U_{i1}$, the vector $U_{i2}$ satisfying (4) is constructed as 
follows :
\begin{eqnarray}
U_{i2} & = & N \epsilon_{ijk} U^{\star}_{j1} U_{l1} ({\cal H}_u){_{lk}} 
\end{eqnarray}
Once $U_{i1}$ and $U_{i2}$ are found, therefore the full matrix $U$ can  
be constructed. \\
Now, since the new $3 \times 3$ hermitian matrices have vanishing (2,1), i.e. 
\begin{eqnarray*} 
( H_{u,d} ){_{21}} & = & 
( T_{u,d} ){_{2i}} (T^{\dagger}_{u,d} ){_{i1}}~~=~~0 
\end{eqnarray*}
it follows that, 
\begin{eqnarray*}
( t_{u,d} ){_{21}} ( t^{\star}_{u,d} ){_{11}} & = & 0
\end{eqnarray*}
where we have used the fact that $T_{u,d}$ is lower triangular. \\
From the nonsingularity of $T_{u,d}$, it implies that :
\begin{eqnarray}
( t_{u,d} ){_{21}} & = & 0
\end{eqnarray}
This completes the proof.
~\\ 
~\\
\underline{ Theorem 2 :} ( for two upper triangular mass matrices ) \\
Given any two nonsingular $3 \times 3$ upper triangular mass matrices 
${\cal T}_u, {\cal T}_d$, there is always a weak 
basis transformation such that the new 
mass matrices $T_u, T_d$ are upper triangular with the matrix elements 
satisfying,  $(t_{u,d}){_{12}} (t^{\star}_{u,d}){_{22}} + 
(t_{u,d}){_{13}} (t^{\star}_{u,d}){_{23}} = 0$. \\
~\\
\underline{Proof :} \\
It is always possible to choose the transformed mass matrices $T_{u,d}$ 
upper triangular with vanishing matrix element $(H_{u,d}){_{12}}$ for 
the hermitian matrix, see above. \\
Now, the requirement $(H_{u,d}){_{12}} = 0$ leads to :
\begin{eqnarray*}
( T_{u,d} ){_{1i}} ( T^{\dagger}_{u,d} ){_{i2}} = 0 
\end{eqnarray*}
and since $T_{u,d}$ are upper triangular, we get :
\begin{eqnarray}
(t_{u,d}){_{12}} (t^{\star}_{u,d}){_{22}} + 
(t_{u,d}){_{13}} (t^{\star}_{u,d}){_{23}} & = & 0 
\end{eqnarray}
End of the proof.
~\\
~\\
Now the following remarks are in order.   
Instead of choosing for both sectors, upper or lower triangular mass matrices, 
we could choose for one sector lower triangular and for the other sector 
upper triangular. In this case, we get for the sector with lower one a 
vanishing matrix element (2,1) and for the upper a relation between 
matrix elements similar to (8). \\
It is clear that we can also find a weak basis such that the lower triangular 
mass matrices $T_u, T_d$ are transformed into lower triangular mass matrices 
with vanishing matrix elements (3,1) or (3,2), 
see Appendix here or ~\cite{ben}. \\
Similarly for upper triangular, the transformed matrix is upper triangular 
with one matrix element written in terms of the others. \\ 
~\\
\section*{Appendix: }
Here, we list all the textures of these new triangular mass matrices 
where we have dropped the flavor indices $u,d$. \\
~\\
{\bf I--weak basis such that $H_{12} = H_{21} = 0$ }, \\
~\\
\underline{ 1--lower triangular : } \\
~\\
a)~~$t_{21} = 0$, \\
\begin{eqnarray*}
T_b & = & \left( \begin{array}{ccc}
t_{11} & 0 & 0 \\
0 & t_{22} & 0 \\
t_{31} & t_{32} & t_{33} \end{array} \right)
\end{eqnarray*}
~\\
b)  singular matrix with $t_{11} = 0$, \\
\begin{eqnarray*}
T_b & = & \left( \begin{array}{ccc}
0 & 0 & 0 \\
t_{21} & t_{22} & 0 \\
t_{31} & t_{32} & t_{33} \end{array} \right)
\end{eqnarray*}
~\\
\underline{ 2--upper triangular : } \\
~\\
\begin{eqnarray*}
T_h & = & \left( \begin{array}{ccc}
t_{11} & - \frac{t_{22} t_{23}^{\star} t_{13}}{|t_{22}|^2} & t_{13} \\
0 & t_{22} & t_{23} \\
0 & 0 & t_{33} \end{array} \right)
\end{eqnarray*}
~\\
{\bf II--weak basis such that $H_{23} = H_{32} = 0$}, \\
~\\
\underline{ 1--lower triangular : } \\
\begin{eqnarray*}
T_b & = & \left( \begin{array}{ccc}
t_{11} & 0 & 0 \\
t_{21} & t_{22} & 0 \\
t_{31} & - \frac{t_{22} t_{21}^{\star} t_{31}}{|t_{22}|^2} & t_{33}  
\end{array} \right)
\end{eqnarray*}
~\\
\underline{2-- upper--triangular matrix :} \\
~\\
a)~~$t_{23} = 0$, \\
\begin{eqnarray*}
T_h & = & \left( \begin{array}{ccc}
t_{11} & t_{12} & t_{13} \\
0 & t_{22} & 0 \\
0 & 0 & t_{33} \end{array} \right)
\end{eqnarray*}
~\\
b) singular matrix with $t_{33} = 0$, \\
\begin{eqnarray*}
T_h & = & \left( \begin{array}{ccc}
t_{11} & t_{12} & t_{13} \\
0 & t_{22} & t_{23} \\
0 & 0 & 0 \end{array} \right)
\end{eqnarray*}
~\\
{\bf III--weak basis such that $H_{13} = H_{31} = 0$ }, \\
~\\
\underline{ 1--lower triangular :} \\
~\\
a) $t_{31} = 0$, \\
\begin{eqnarray*}
T_b & = & \left( \begin{array}{ccc}
t_{11} & 0 & 0 \\
t_{21} & t_{22} & 0 \\
0 & t_{23} & t_{33} \end{array} \right)
\end{eqnarray*}
~\\
b)  singular matrix for $t_{11} = 0$, \\ 
\begin{eqnarray*}
T_b & = & \left( \begin{array}{ccc}
0 & 0 & 0 \\
t_{21} & t_{22} & 0 \\
t_{31} & t_{32} & t_{33} \end{array} \right)
\end{eqnarray*}
~\\
\underline{ 2--upper triangular :} \\ 
~\\
a) $t_{13} = 0$, \\ 
\begin{eqnarray*}
T_h & = & \left( \begin{array}{ccc}
t_{11} & t_{12} & 0 \\
0 & t_{22} & t_{23} \\
0 & 0 & t_{33} \end{array} \right)
\end{eqnarray*} 
~\\
b)  singular matrix for $t_{33} = 0$, \\
\begin{eqnarray*}
T_b & = & \left( \begin{array}{ccc}
t_{11} & t_{12} & t_{13} \\
0 & t_{22} & t_{23} \\
0 & 0 & 0 \end{array} \right)
\end{eqnarray*}
\section*{Acknowlegment :}
I would like to thank the DAAD for its financial 
support. Special thanks go to Prof. Scheck for reading the manuscript 
and giving me his comments.


\begin{thebibliography}{99}
\bibitem{fri}
H. Fritzsch, Phys. Lett. {\bf B73} (1978) p.317 \\
Nucl. Phys. {\bf B155} (1979) p.182
\bibitem{wil}
F. Wilczek, A. Zee,  Phys. Rev. Lett. {\bf 42} (1979) p.421 
\bibitem{geo}
H. Georgi, C. Jarlskog, Phys. Lett. {\bf B86} (1979) p.297  
\bibitem{ram}
P. Ramond, R. G. Roberts, G.G. Ross, Nucl. Phys. {\bf B406} (1993) p.19
\bibitem{coq}
R. Coquereaux, G. Esposito--Far\`ese, F. Scheck, Int. J. Mod. Phys. 
{\bf A7} (1992) p.6555 \\  
R. H\"au{\ss}ling, N.A. Papadopoulos, F. Scheck, Phys. Lett. 
{\bf B260} (1991) 
p.125 \\
R. Coquereaux, R. H\"au{\ss}ling, N.A. Papadopoulos, F. Scheck, 
Int. J. Mod. Phys. {\bf A7} (1992) p.2809 
\bibitem{hau}
R. H\"au{\ss}ling, F. Scheck, Phys. Lett. {\bf B336} (1994) p.477 
\bibitem{bra}
Branco et al., Phys. Rev. {\bf D39} (1989) p.3443 
\bibitem{ben}
H.B. Benaoum, F. Scheck, " New textures for triangular mass matrices " 
Mainz--preprint, {\bf MZ--TH/98--36}  
\bibitem{sss}
H.B. Benaoum, R. H\"au{\ss}ling and F. Scheck, work in preparation. 
\bibitem{sch}
R. H\"au{\ss}ling,  F. Scheck,  Phys. Rev. {\bf D57 } (1998) p.6656 
\end{thebibliography}
\end{document}